\documentstyle[12pt]{article}

\begin{document}
\title{\bf {Casimir densities for parallel plate in the Domain Wall background }}
\author{M.R. Setare  \footnote{E-mail: rezakord@yahoo.com}
  \\
{Physics Dept. Inst. for Studies in Theo. Physics and
Mathematics(IPM)}\\
{P. O. Box 19395-5531, Tehran, IRAN }\\
and \\ {Department of Science, Physics group, Kordestan
University, Sanandeg,  Iran }\\and \\{ Department of Physics,
Sharif University of Technology, Tehran, Iran }}

\date{\small{\today}}
 \maketitle

\begin{abstract}
The Casimir forces on two parallel plates in conformally flat
domain wall background due to conformally coupled massless scalar
field satisfying mixed boundary conditions on the plates is
investigated. In the general case of mixed boundary conditions
formulae are derived for the vacuum expectation values of the
energy-momentum tensor and vacuum forces acting on boundaries.
 \end{abstract}
\newpage

 \section{Introduction}
   The Casimir effect is one of the most interesting manifestations
  of nontrivial properties of the vacuum state in quantum field
  theory [1,2]. Since its first prediction by
  Casimir in 1948\cite{Casimir} this effect has been investigated for
  different fields having different boundary geometries[3-7]. The
  Casimir effect can be viewed as the polarization of
  vacuum by boundary conditions or geometry. Therefore, vacuum
  polarization induced by a gravitational field is also considered as
  Casimir effect.\\
  Casimir effect may  have interesting implications for the early
  universe, in \cite{set3} the
   Casimir effect of a massless scalar field with Dirichlet
   boundary condition in spherical shell having different vacuua
   inside and outside which represents a bubble in early universe with
   false/true vacuum inside/outside have been investigated. The Casimir stress on two
    concentric spherical
   shell with constant comoving radius having different vacuua inside and
    outside in de Sitter space which is corresponding to a
    spherical symmetry domain wall with thickness have been calculated in \cite{set4}.\\
  In the context of hot big bang cosmology, the unified theories
  of the fundamental interactions predict that the universe passes
  through a sequence of phase transitions. Different types of topological objects may
  have been formed during these phase transitions, these include domain walls, cosmic
strings and monopoles \cite{{zel},{kib},{viel}}.These topological
defects appear as a consequence of breakdown of local or global
gauge symmetries of a system composed by self-coupling iso-scalar
Higgs fields
$\Phi^a$.\\
 It has been shown in \cite{Vilenkin1} and
\cite{Vilenkin2}, that the gravitational field of the vacuum
domain wall with a source of the form
\begin{equation}
 T_\mu^\nu=\sigma \delta (x) diag(1,\, 0,\, 1,\, 1)
\label{wallemt}
 \end{equation}
 does not correspond to any exact static
solution of Einstein equations (on domain wall solutions of
Einstein-scalar-field equations see \cite{Widrow}). However the
static solutions can be constructed in presence of an additional
background energy-momentum tensor. Such a type solution has been
fond in \cite{Mansouri}.\\
In the present paper we will investigate the vacuum expectation
values of the energy--momentum tensor of the conformally coupled
scalar field on background of the conformally flat domain wall
geometries. We will consider the general plane--symmetric
solutions of the gravitational field equations and boundary
conditions of the Robin type on the plates. The latter includes
the Dirichlet and Neumann boundary conditions as special cases.
The Casimir energy-momentum tensor for these geometries can be
generated from the corresponding flat spacetime results by using
the standard transformation formula. Previously this method has
been used in \cite{{Seta01},{set2}} to derive the vacuum
characteristics of the Casimir configuration on background of the
static domain wall geometry for a scalar field with Dirichlet
boundary condition on plates. Also this method has been used in
\cite{set6} to derive the vacuum characteristics of the Casimir
configuration on background of conformally flat brane-world
geometries for massless scalar field with Robin boundary condition
on plates. For Neumann or more general mixed boundary conditions
we need to have the Casimir energy-momentum tensor for the flat
spacetime counterpart in the case of the Robin boundary conditions
with coefficients related to the metric components of the domain
wall geometry. The Casimir effect for the general Robin boundary
conditions on background of the Minkowski spacetime was
investigated in Ref. \cite{RomSah} for flat boundaries, here we
use the results of Ref. \cite{RomSah} to generate vacuum
energy--momentum tensor for the conformally flat background.

\section{Vacuum expectation values for the energy-momentum tensor}
In this paper we shall consider the conformally coupled real
scalar field $ \phi $, which satisfies
\begin{equation}
( \Box +  \frac {1} {6} R ) \phi = 0 , \qquad
  \Box =  \frac {1} {\sqrt{ -g }} \partial_{\mu}
( \sqrt{ -g } g^{ \mu \nu } \partial_{\nu} ),  \label{motioneq}
\end{equation}
and propagates on background of gravitational field generated by
domain wall solution from \cite{Vilenkin1}. In \cite{Vilenkin1}
Vilenkin has found the gravitational of domain walls in the linear
approximation of general relativity. At first we review this work
briefly. In linear approximation the metric is given by
\begin{equation}
g_{\mu\nu}=\eta_{\mu\nu}+h_{\mu\nu}, \label{lineq}
\end{equation}
where $\eta_{\mu\nu}=diag(1,-1,-1,-1)$, and $|h_{\mu\nu}|\ll 1$,
Einstein equations are as following
\begin{equation}
( \nabla^{2}-\partial _{t}^{2}) h_{\mu\nu}=16\pi G
(T_{\mu\nu}-1/2\eta_{\mu\nu}T), \label{eineq}
\end{equation}
with the harmonic coordinate conditions
\begin{equation}
\partial_{\nu}(h_{\mu}^{\nu}-1/2\delta_{\mu}^{\nu}h)=0. \label{hareq}
\end{equation}
The remaining coordinate freedom is restricted to the
transformations
\begin{equation}
h'_{\mu\nu}=h_{\mu\nu}-\xi_{\mu,\nu}-\xi_{\nu,\mu} . \label{freq}
\end{equation}
where
\begin{equation}
( \nabla^{2}-\partial _{t}^{2})\xi_{\mu}=0 . \label{freq1}
\end{equation}
The solution of Eqs. (\ref{eineq}) and (\ref{hareq}) for a vacuum
domain wall with energy-momentum tensor given by
Eq.(\ref{wallemt}) (For a vacuum domain wall  we have $-p=\sigma$
where $-p$ is the surface tension and $\sigma$ is the surface
energy density) is easily found where
\begin{equation}
h_{00}=-h_{22}=-h_{33}=-4\pi G \sigma|x|, \hspace{1cm}
h_{11}=12\pi G \sigma|x| . \label{soleq}
\end{equation}
If we consider the coordinate transformation Eq.(\ref{freq}) with
$\xi_{1}=2\pi G \sigma x^{2} sgn x$ and $\xi_2=\xi_3=\xi_0=0$
brings the metric Eq.(\ref{soleq}) to a conformally flat
form
\begin{equation} ds^2=(1-4\pi G \sigma |x|)( dt^2- dx^2- dy^2
- dz^2). \label{metric}
\end{equation}
In Eq. (\ref{motioneq}) $R$ is the Ricci scalar for the metric
$g_{\mu \nu }$. Note that for the metric tensor from Eq.
(\ref{metric}) one has
\begin{equation}
R=-4\pi G\sigma\frac{-(sgn(x))'+4(sgn(x))'\pi G\sigma|x|-2\pi
G\sigma(sgn(x))^{2}}{(1-4\pi G\sigma|x|)^{3}}, \label{Riccisc}
\end{equation}
In what follows as a boundary configuration we shall consider two
plates parallel to each other and to domain wall, with $x$
coordinates equal to $x_1$ and $x_2$ (to be definit we shall
consider right half space of domain wall geometry $ x_1 , x_2 > 0
$).We will assume that the field satisfies the mixed boundary
condition
\begin{equation}
\left( a_{j}+b_{j}n^{\mu }\nabla _{\mu }\right) \varphi
(x)=0,\quad x=x_{j},\quad j=1,2  \label{boundcond}
\end{equation}
on the plate $x=x_{1}$ and $x=x_{2}$, $x_{1}<x_{2}$, $n^{\mu }$ is
the normal to these surfaces, $n_{\mu }n^{\mu }=-1$, and $a_j$,
$b_j$ are constants. The results in the following will depend on
the ratio of these coefficients only. However, to keep the
transition to the Dirichlet and Neumann cases transparent we will
use the form (\ref{boundcond}). For the case of plane boundaries
under consideration introducing a new coordinate $u$ in accordance
with
\begin{equation}
du=\sqrt{1-4\pi G \sigma |x|}dx  \label{ycoord}
\end{equation}
conditions (\ref{boundcond}) take the form
\begin{eqnarray}\label{boundcond1}
\left( a_{j}+(-1)^{j-1}b_{j}\frac{1}{\sqrt{1-4\pi G \sigma
|x|}}\partial _{x}\right) \varphi (x) &=&\\
 \left(
a_{j}+(-1)^{j-1}b_{j}\partial _{u}\right) \varphi (x)&=&0,\quad
u=u_{j},\quad j=1,2.\nonumber
\end{eqnarray}
Note that the Dirichlet and Neumann boundary conditions are
obtained from Eq. (\ref{boundcond}) as special cases corresponding
to $(a_j,b_j)=(1,0)$ and $(a_j,b_j)=(0,1)$ respectively. The Robin
boundary condition may be interpreted as the boundary condition on
a thick plate \cite{leb}. Rewriting Eq.(\ref{boundcond1}) in the
following form
\begin{equation}
\varphi(x)=-(-1)^{j-1}\frac{b_{j}}{a_{j}}\partial u\varphi(x),
\end{equation}
where $\frac{a_{j}}{b_{j}}$, having the dimension of a length, may
be called skin-depth parameter. This is similar to the case of
penetration of an electromagnetic field into a real metal, where
the tangential component of the electric
field is proportional to the skin-depth parameter.\\
 Our main interest in the present paper is to investigate the vacuum
expectation values (VEV's) of the energy--momentum tensor for the field $%
\varphi (x)$ in the region $x_{1}<x<x_{2}$. The presence of
boundaries modifies the spectrum of the zero--point fluctuations
compared to the case without boundaries. This results in the shift
in the VEV's of the physical quantities, such as vacuum energy
density and stresses. This is the well known Casimir effect. It
can be shown that for a conformally coupled scalar by using field
equation (\ref{motioneq}) the expression for the energy--momentum
tensor can be presented in the form
\begin{equation}
T_{\mu \nu }=\nabla _{\mu }\varphi \nabla _{\nu }\varphi -\frac{1}{6} \left[ \frac{%
g_{\mu \nu }}{2}\nabla _{\rho }\nabla ^{\rho }+\nabla _{\mu
}\nabla _{\nu }+R_{\mu \nu }\right] \varphi ^{2},  \label{EMT1}
\end{equation}
where $R_{\mu \nu }$ is the Ricci tensor. The quantization of a
scalar filed on background of metric Eq.(\ref{metric}) is
standard. Let $\{\varphi _{\alpha }(x),\varphi _{\alpha }^{\ast
}(x)\}$ be a complete set of orthonormalized positive and negative
frequency solutions to the field equation (\ref {motioneq}),
obying boundary condition (\ref{boundcond}). By expanding the
field operator over these eigenfunctions, using the standard
commutation rules and the definition of the vacuum state for the
vacuum expectation values of the energy-momentum tensor one
obtains
\begin{equation}
\langle 0|T_{\mu \nu }(x)|0\rangle =\sum_{\alpha }T_{\mu \nu }\{\varphi {%
_{\alpha },\varphi _{\alpha }^{\ast }\}},  \label{emtvev1}
\end{equation}
where $|0\rangle $ is the amplitude for the corresponding vacuum
state, and the bilinear form $T_{\mu \nu }\{{\varphi ,\psi \}}$ on
the right is determined by the classical energy-momentum tensor
(\ref{EMT1}). In the problem under consideration we have a
conformally trivial situation: conformally invariant field on
background of the conformally flat spacetime. Instead of
evaluating Eq. (\ref{emtvev1}) directly on background of the
curved metric, the vacuum expectation values can be obtained from
the corresponding flat spacetime results for a scalar field
$\bar{\varphi}$ by using the conformal properties of the problem
under consideration. Under the
conformal transformation $g_{\mu \nu }=\Omega ^{2}\eta _{\mu \nu }$ the $%
\bar{\varphi}$ field will change by the rule
\begin{equation}
\varphi (x)=\Omega ^{-1}\bar{\varphi}(x),  \label{phicontr}
\end{equation}
where for metric Eq.(\ref{metric}) the conformal factor is given
by $\Omega =\sqrt{1-4\pi G \sigma |x|}$. The boundary conditions
for the field $\bar{\varphi}(x)$ we will write in form similar to
Eq. (\ref{boundcond1})
\begin{equation}
\left( \bar{a}_{j}+(-1)^{j-1}\bar{b}_{j}\partial _{x}\right) \bar{\varphi}%
=0,\quad x=x_{j},\quad j=1,2,  \label{bounconflat}
\end{equation}
with constant Robin coefficients $\bar{a}_{j}$ and $\bar{b}_{j}$.
Comparing to the boundary conditions (\ref{boundcond}) and taking
into account transformation rule (\ref{phicontr}) we obtain the
following relations between the corresponding Robin coefficients
\begin{equation}
\bar{a}_{j}=a_{j}+(-1)^{j-1}\frac{2\pi G\sigma sgn(x) }{(1-4\pi
G\sigma |x|)^{3/2} }b_{j},\quad
\bar{b}_{j}=b_{j}\frac{1}{\sqrt{1-4\pi G\sigma
|x|}}\label{coefrel}
\end{equation}
Note that as Dirichlet boundary conditions are conformally
invariant the Dirichlet scalar in the curved bulk corresponds to
the Dirichlet scalar in a flat spacetime. However, for the case of
Neumann scalar the flat spacetime counterpart is a Robin scalar
with
\begin{equation}
\bar{a}_j=(-1)^{j-1}\frac{2\pi G\sigma sgn(x)}{1-4\pi G\sigma
|x|}, \hspace{1cm} \bar{b}_j=1 \label{abeq}
\end{equation}
The Casimir effect with boundary conditions (\ref{bounconflat}) on
two parallel plates on background of the Minkowski spacetime is
investigated in Ref. \cite{RomSah} for a scalar field with a
general conformal coupling parameter. In the case of a conformally
coupled scalar the corresponding regularized VEV's for the
energy-momentum tensor are uniform in the region between the
plates and have the form
\begin{equation}
\langle \bar{T}_{\nu }^{\mu }\left[ \eta _{\alpha \beta }\right] \rangle _{%
{\rm ren}}=-\frac{J_3(B_1,B_2)}{8\pi ^{3/2}a^{4}\Gamma (5/2)}{\rm
diag}(1,1,1,-3), \quad x_{1}< x< x_{2}, \label{emtvevflat}
\end{equation}
where
\begin{equation}\label{IDB1B2}
  J_3(B_1,B_2)={\rm p.v.}
\int_{0}^{\infty }\frac{t^{3}dt}{\frac{(B_{1}t-1)(B_{2}t-
1)}{(B_{1}t+1)(B_{2}t+1)}e^{2t}-1},
\end{equation}
and we use the notations
\begin{equation}
B_{j}=\frac{\bar{b}_{j}}{\bar{a}_{j}a},\quad j=1,2,\quad
a=x_{2}-x_{1}. \label{Bjcoef}
\end{equation}
For the Dirichlet scalar $B_1=B_2=0$ and one has
$J_D(0,0)=2^{-4}\Gamma (4)\zeta _R(4)$, with the Riemann zeta
function $\zeta _R(s)$. Note that in the regions $x< x_{1}$ and
$x> x_{2}$ the Casimir densities vanish \cite{{Seta01},{RomSah}}:
\begin{equation}
\langle \bar{T}_{\nu }^{\mu }\left[ \eta _{\alpha \beta }\right] \rangle _{%
{\rm ren}}=0,\quad x< x_{1},x> x_{2}.  \label{emtvevflat2}
\end{equation}
This can be also obtained directly from Eq. (\ref{emtvevflat})
taking the limits $x_{1}\rightarrow -\infty $ or $x_{2}\rightarrow
+\infty $. The values of the coefficients $B_{1}$ and $B_{2}$ for
which the denominator in the subintegrand of Eq.
(\ref{emtvevflat}) has zeros are specified in \cite{RomSah}. The
vacuum energy-momentum tensor on domain wall background
Eq.(\ref{metric}) is obtained by the standard transformation law
between conformally related problems (see, for instance,
\cite{davies}) and has the form
\begin{equation}
\langle T_{\nu }^{\mu }\left[ g_{\alpha \beta }\right] \rangle _{{\rm ren}%
}=\langle T_{\nu }^{\mu }\left[ g_{\alpha \beta }\right] \rangle _{{\rm ren}%
}^{(0)}+\langle T_{\nu }^{\mu }\left[ g_{\alpha \beta }\right] \rangle _{%
{\rm ren}}^{(b)}.  \label{emtcurved1}
\end{equation}
Here the first term on the right is the vacuum energy--momentum
tensor for the situation without boundaries (gravitational part),
and the second one is due to the presence of boundaries. As the
quantum field is conformally coupled and the background spacetime
is conformally flat the gravitational part of the energy--momentum
tensor is completely determined by the trace anomaly and is
related to the divergent part of the corresponding effective
action by the relation \cite{davies}
\begin{equation}
\langle T_{\nu }^{\mu }\left[ g_{\alpha \beta }\right] \rangle _{{\rm ren}%
}^{(0)}=2g^{\mu \sigma }(x)\frac{\delta }{\delta g^{\nu \sigma }(x)}W_{{\rm %
div}}[g_{\alpha \beta }].  \label{gravemt}
\end{equation}
The boundary part in Eq. (\ref{emtcurved1}) is related to the
corresponding flat spacetime counterpart
(\ref{emtvevflat}),(\ref{emtvevflat2}) by the relation
\cite{davies}
\begin{equation}
\langle T_{\nu }^{\mu }\left[ g_{\alpha \beta }\right] \rangle _{{\rm ren}%
}^{(b)}=\frac{1}{\sqrt{|g|}}\langle \bar{T}_{\nu }^{\mu }\left[
\eta _{\alpha \beta }\right] \rangle _{{\rm ren}}.
\label{translaw}
\end{equation}
By taking into account Eq. (\ref{emtvevflat}) from here we obtain
\begin{equation}
\langle T_{\nu }^{\mu }\left[ g_{\alpha \beta }\right] \rangle _{{\rm ren}%
}^{(b)}=-\frac{J_3(B_1,B_2)}{8\pi ^{3/2}a^{4}\Gamma (5/2)(1-4\pi G \sigma |x|)^{2}}%
{\rm diag}(1,1,1,-3),  \label{bpartemt}
\end{equation}
for $x_{1}< x< x_{2}$, and
\begin{equation}
\langle T_{\nu }^{\mu }\left[ g_{\alpha \beta }\right] \rangle _{{\rm ren}%
}^{(b)}=0,\;{\rm for}\;x< x_{1},x> x_{2}.  \label{bpartemt2}
\end{equation}
In Eq. (\ref{bpartemt}) the constants $B_{j}$ are related to the
Robin coefficients in boundary condition (\ref{boundcond}) by
formulae (\ref {Bjcoef}),(\ref{coefrel}) and are functions on
$x_j$. In particular, for Neumann boundary conditions
$B^{(N)}_j=\frac{(-1)^{j-1}}{a}\frac{1-4\pi G\sigma |x|}{2\pi G
\sigma sgn(x)}$.\\
The total bulk vacuum energy per unit physical surface on the
plates at $x=x_j$ is obtained by integrating over the region
between the plates
\begin{equation}\label{bulktoten}
  E_j^{(b)}=(1-4\pi G\sigma |x_{j}|)^{-3/2} \int _{x_1}^{x_2}\langle T_0^0\rangle
  ^{(b)}_{{\mathrm{ren}}}(1-4\pi G\sigma |x|)^{2}dx=-\frac{J_3(B_1,B_2)
  (1-4\pi G\sigma |x_{j}|)^{-3/2}}{8\pi ^{3/2}\Gamma (5/2)a^3}.
\end{equation}
The resulting vacuum force per unit boundary area acting on the
boundary at $x=x_{j}$ is determined by the difference
\begin{equation}
\langle T_{3}^{3}\left[ g_{\alpha \beta }\right] \rangle _{{\rm ren}%
}|_{x=x_{j}+0}-\langle T_{3}^{3}\left[ g_{\alpha \beta }\right] \rangle _{%
{\rm ren}}|_{x=x_{j}-0}.  \label{vacforce0}
\end{equation}
 The first term in Eq.(19) is the vacuum
polarization due to the gravitational field, without any boundary
conditions, which
 can be rewritten as following
\begin{equation}
\langle T_{\nu }^{\mu }\left[ g_{\alpha \beta }\right] \rangle _{{\rm ren}%
}^{(0)}=-\frac{1}{2880}[\frac{1}{6} \tilde H^{(1)\mu}_{\nu}-\tilde
H^{(3)\mu}_{\nu}]. \label{heq}
\end{equation}
 The
functions $H^{(1,3)\mu}_{\nu}$ are some combinations of curvature
tensor components (see \cite{davies}). Now we see that as
gravitational part (\ref{heq}) is a continous function on $x$ it
does not contribute to the forces acting on the
boundaries and the vacuum force per unit surface acting on the boundary at $%
x=x_{j}$ is determined by the boundary part of the vacuum pressure
, $%
p_{b}=-\langle T_{3}^{3}\left[ g_{\alpha \beta }\right] \rangle _{{\rm ren}%
}^{(b)}$, taken at the point $x=x_{j}$:
\begin{equation}
p_{bj}(x_{1},x_{2})=-\frac{(1-4\pi G \sigma |x_{j}|) ^{2}
J_3(B_1,B_2)}{4\pi ^{3/2}a^{4}\Gamma (3/2)} . \label{vacforce}
\end{equation}
This corresponds to the attractive/repulsive force between the plates if $%
p_{bj}</>0$. The equilibrium points for the plates correspond to
the zero values of Eq. (\ref{vacforce}): $p_{bj}=0$. These points
are zeros of the function $J_3(B_1,B_2)$ defined by Eq.
(\ref{IDB1B2}) and are the same for both plates. Hence, we have an
example for the stabilization of the distance between the plates
due to the vacuum forces (To see an application to the
Randall-Sundrum brane-world model refer to \cite{set6}). Note that
at these points the VEV's of the bulk energy-momentum tensor given
by Eq. (\ref{bpartemt}) and the total bulk energy also vanish.

 \section{Conclusion}
 The study of vacuum quantum effect on the background of
 topological defects such as domain walls, cosmic strings, or
 magnetic monopoles have been considered in many references
 \cite{{ander},{kho},{Hisc90},{Mazz91},{Beze99},{Bord96m},
 {Beze00},{sast1},{sast2}}.
In the present paper we have investigated the Casimir effect for a
conformally coupled massless scalar field confined in the region
between two parallel plate  on background of the conformally-flat
domain wall. The general case of the mixed(Robin) boundary
conditions is considered. The vacuum expectation values of the
energy-momentum tensor are derived from the corresponding flat
spacetime results by using the conformal properties of the
problem. This method has been used in \cite{set6} to derive the
vacuum characteristics of the Casimir configuration on background
of conformally flat brane-world geometries for massless scalar
field with Robin boundary condition on plates, then as an
application of the general formulae, there we have considered the
important special case of the AdS$_{5}$ bulk, in odd spacetime
dimensions the conformal anomaly is absent and the corresponding
gravitational vacuum polarization vanishes. In the present paper,
spacetime have even dimension, the vacuum polarization due to the
gravitational field, without any boundary conditions is not zero
and given by Eq.(\ref{heq}), the corresponding gravitational
pressure part has the same from both sides of the plates, and
hence leads to zero effective force.
 In the region between the plates the boundary induced
part for the vacuum energy-momentum tensor is given by
Eq.(\ref{bpartemt}), and the corresponding vacuum forces acting
per unit surface of the plates have the form Eq. (\ref{vacforce}).
These forces vanish at the zeros of the function $J_3(B_1,B_2)$.

\end{document}